# Setup for measuring the optical propagation time in matter

Shenghao Wang, Shijie Liu[*], You Zhou, Jianda Shao

Testing center, Shanghai Institute of Optics and Fine Mechanics, Chinese Academy of Sciences, Shanghai, 201800, China.

* Corresponding author: shijieliu@siom.ac.cn

**Abstract:** Optical propagation time in matter could reveal fruitful information, such as the velocity of light and the sample's refractive index. In this paper, we build a simple and robust setup for measuring the optical propagation time in matter for a known distance, the system uses high frequency square signal as the signal carrier, and a lock-in amplifier is employed to obtain the phase difference between the reference square signal and the other one penetrating the sample, in this way the optical time of flight in matter can be obtained by a background subtraction process. Primary experimental result confirms the feasibility of the newly proposed measuring theory, which can be used to measure easily in high-speed the velocity of light and the refractive index of optical transparent material, compared with the currently popular measuring technique using oscilloscope, potential advantage of our proposed method employing lock-in amplifier is that high accuracy are promising, and in contrast with the presently most popular method for determining the sample's refractive index based on the minimum deviation angle, superiority of our suggested method is the easy preparation of the sample, the convenient operability and the fast measuring speed.

## 1. Introduction

Measurement of the optical propagation time in matter at a known distance could be used to obtain the velocity of light, which thus plays an important role in the history of science. Early in the year of 1638, Galileo Galilei has attempted to measure the time of flight between two mountains by watching the lighted lanterns with the naked eye [1], later respectively in 1849 and 1862, Hippolyte Fizeau and Léon Foucault successively measured the optical propagation time at a known distance by ingeniously associating it with the rotation speed of cogwheel and mirror [2-4], and a relative measurement error of about 5% existed. In order to decrease the relative error of recording the optical time of flight, in 1926, Albert Abraham Michelson carried out the measurement of optical propagation time following the above manner between Mount Wilson and Mount San Antonio, some 22 miles distant, and a high accuracy of 0.003% was achieved mainly resulting from the extremely long propagation distance [5].

However, it is clear that the aforementioned technique of measuring the optical propagation time is incredibly inflexible and difficult to implement in laboratory, nowadays, using oscilloscopes with time resolutions of less than one nanosecond, the optical propagation time can be directly measured by timing the delay of a light pulse from a laser or an LED reflected from a mirror, this method is less precise (with errors in the order of 1%), but it is very easy to carry out in a common optical laboratory, and it has found wide application as a laboratory experiment in college physics classes [6-15].

In this manuscript, different from the currently popular measuring technique using oscilloscopes, we build a simple setup for measuring the optical time of flight in air for a known distance employing a lock-in amplifier, the system use high frequency square signal as the signal carrier, and a lock-in amplifier is employed to obtain the phase difference between the reference square signal and the other one penetrating the sample, in this way the time of flight in air can be obtained by a background subtraction process.

The following sections will be arranged like this, firstly we will describe layout and working principle of the system for measuring the optical propagation time in air, and then a primary experiment verifying the method's feasibility will be demonstrated, after that, we will provide implementation process using this facility to measure the velocity of light and the refractive index of optical transparent materials.

## 2. Materials and methods

### 2.1 Experimental setup

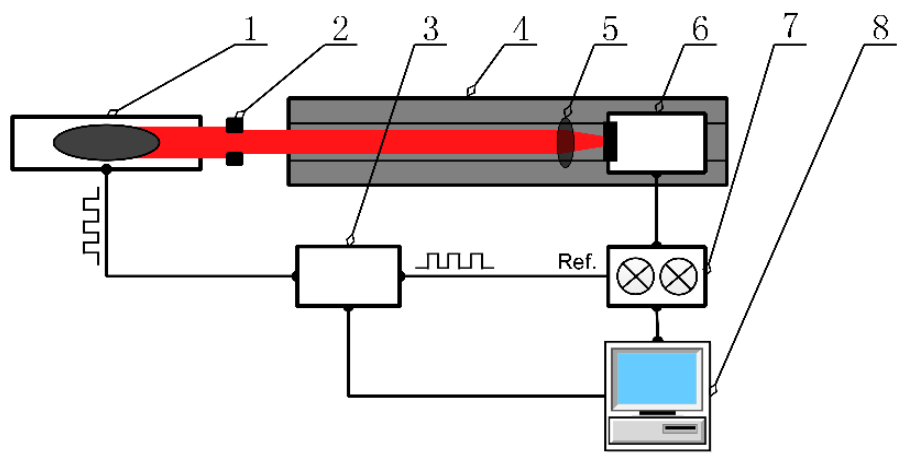

Fig. 1 Framework of the system for measuring the optical propagation time in air, 1- laser diode module, 2- aperture, 3- function generator, 4- linear stage, 5- focusing lens, 6- high speed detector, 7- lock-in amplifier, 8- personal computer.

As shown in Fig. 1 is the framework of the system for measuring the optical propagation time in matter, it is mainly made up of a laser diode module, an aperture, a function generator, a linear stage, a focusing lens, a high speed photo detector, a lock-in amplifier and a personal computer. The laser diode module (Manufactory: Newport, Model: LQA1064-150E), which can be analog modulated up to 20 MHz by an external voltage signal, emits an elliptical beam with a total power of 150 mW, and the center wavelength of its output spectrum is 1064 nm with ±15 nm accuracy. The function generator (Manufactory: Tektronix, Model: AFG3022C) can produce various kinds of waveforms through two output channels with a maximum frequency of 20 MHz. The linear stage (Manufactory: Shanghai Lian Yi, Model: ORD 100) can move with high straightness by a controller (Manufactory: Shanghai Lian Yi, Model: KCP-1). The focusing lens (Manufactory: Thorlabs, Model: LA1027-ML) works in the wave range from 350 nm to 2000 nm, and its focal length is 35 mm at the wavelength of 633 nm. The high speed photo detector (Manufactory: Newport, Model: 1811-FS) is an InGaAs optical receiver with a 3 dB bandwidth of 125 MHz, its spectral response wave range is from 900 nm to 1700 nm and the rise time is 3 ns. The lock-in amplifier (Manufactory:

Signal Recovery, Model: 7270) can realize all measurements over a frequency range extending from 1 mHz to 250 kHz with a sensitivity of 2 nV.

**2.2 Working principle and measuring theory**

Based on the optical setup demonstrated in Fig. 1, the measuring theory of the optical propagation time in air for a certain distance is described as follows. The function generator works here to produce square waveform for modulating the laser diode through one output channel and meanwhile provides a reference square signal with the same frequency via the other channel for the lock-in amplifier. The laser diode module, under the action of the square waveform released from the function generator, firstly emits off and on an elliptical light beam in the same frequency with the input square waveform, and then after shaped by the aperture, the generated circular light beam penetrates through the air and is finally collected by the high speed photo detector after focused by the focusing lens, the photo signal collected by the detector is firstly photovoltaic transformed and then the output electrical signal is captured and digitized by the lock-in amplifier, which is employed to obtain the phase difference between the reference square signal produced by the function generator and the electrical signal emitted from the high speed photo receiver (the working principle of lock-in amplifier for measuring the phase difference between two input signals can be referred at literature [16, 17]). The relative position between the detector and laser source can be adjusted by the linear motorized stage. The personal computer is used to achieve remote control of the function generator and the lock-in amplifier through USB interfaces, a LabVIEW-based software was developed to implement the desired functionality such as system initialization, moving the linear stage, frequency setting & scanning, data acquisition, data post processing, graphic display and data storage.

Suppose the square waveform released from the function generator transferred respectively to the laser diode module and the lock-in amplifier is $f$, and $\theta_0$ represents the phase difference measured by the lock-in amplifier when the detector is in the initial position, while $\theta_1$ expresses the phase difference after moving simultaneously the focusing lens and the detector by the linear stage along the beam path for a known distance $d$, then the optical propagation time in air for the distance of $d$ can be written as:

$$t = \frac{|\theta_1 - \theta_0|}{360} \times \frac{1}{f} \tag{1.1}$$

Here $\theta_0$ and $\theta_1$ are recorded in the unit of degree.

## 3. Experimental result and discussion

Our primary experiment was carried out without adjusting the relative position between the photo detector and the laser source, and the data was acquired in the following process.

S1: Set the frequency of the square waveform emitted through the two output channels of the function generator both to 1 kHz, and after aligning their initial phase, use the lock-in amplifier to obtain the phase difference between the reference square signal produced by the function generator and the electrical signal emitted from the high speed photo receiver.

S2: In the step of 1 kHz, successively set the frequency of the generated square waveform to 2 kHz, 3 kHz, 4 kHz $\cdots$ 99 kHz and 100 kHz, and repeat S1 at each frequency to obtain the phase difference.

S3: Plot the curve of phase difference versus frequency.

Theoretically speaking, suppose the time difference between the two square signals when reaching the lock-in amplifier (one square signal transfers directly from the function generator to the lock-in amplifier, while the other experiences the following procedure: modulating the laser diode module → propagating through the air → responded by the high-speed detector → photovoltaic transformed → transferring to the lock-in amplifier) is a constant $\Delta t$, and then the frequency dependence relationship of the phase difference $\Delta\theta$ between the two square signal can be written as:

$$\Delta\theta = \frac{\Delta t}{360} \times f \tag{1.2}$$

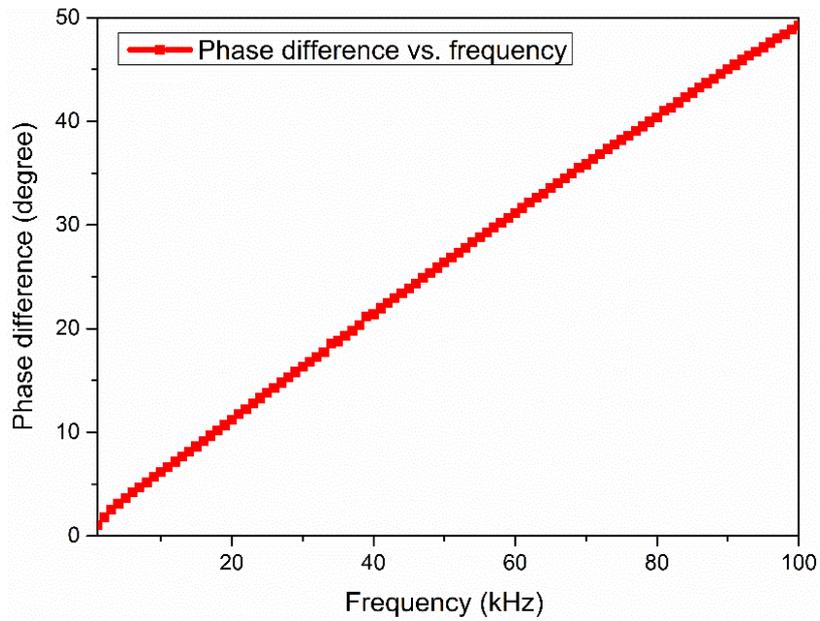

Fig. 2 The measured curve of phase difference versus frequency.

As shown in Fig. 2 is the measured curve of phase difference versus frequency, and we can see clearly from Fig. 2 that the phase difference shows a good linear relationship versus the frequency, which confirms our aforementioned theoretical analysis. Through a linear fitting, the calculated slope of the curve is 4.858 ms and the time difference between the two square signals when reaching the lock-in amplifier is 0.1749 s.

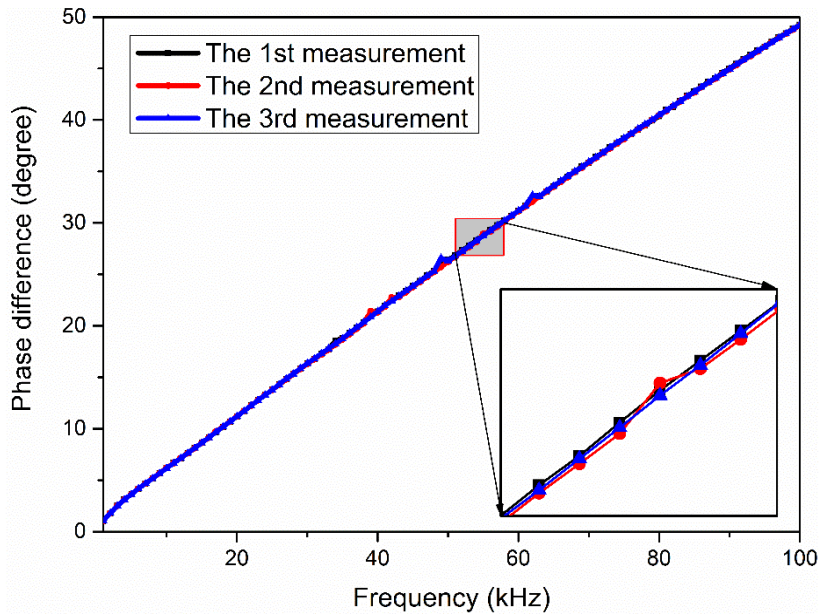

Fig. 3 Three times repetitive measurement of the curves of phase difference versus frequency.

Fig. 3 represents the three times repetitive experiment result for the curve of phase difference versus frequency, and the repeatability error of the three times repetitive experiment can be written as:

$$\zeta = \frac{\sum_{k=1}^{100} \chi_{f_k}}{100} = 0.159\% \quad (1.3)$$

Where $\chi_{f_k}$ is the repeatability error at the frequency of $f_k$, and it was computed as follow:

$$\chi_{f_k} = \frac{\sqrt{\frac{\sum_{i=1}^{3}\left(\theta_i - \frac{\sum_{i=1}^{3}\theta_i}{3}\right)^2}{3}}}{\frac{\sum_{i=1}^{3}\theta_i}{3}} \quad (1.4)$$

In equation (1.4), $\theta_i$ represents the phase difference at frequency of $f_k$ in the $i$ th measurement.

Through linear fitting, the obtained time difference of the three times measurement are respectively 0.1749 s, 0.1745 s and 0.1749 s. Combing the contrast curves (see the enlarged view at the bottom right corner) as shown in Fig. 3 and the aforementioned computed repeatability error (0.159%), it can be concluded that the measuring strategy has very good repeatability accuracy.

## 4. Applications of this facility

### 4.1 Light velocity measurement

Using the aforementioned optical setup demonstrated in Fig. 1, the measurement of light velocity in air can be realized in the following manner.

S1: Set the frequency of the square waveform emitted through the two output channels of the function generator both to $f$, then use the lock-in amplifier to obtain the phase difference $\Delta\theta_0$ between the reference square signal produced by the function generator and the electrical signal emitted from the high speed photo receiver.

S2: Move simultaneously the focusing lens and the high-speed detector by the linear stage for a distance of $d$ along the beam path, then repeat S1, and write the obtained phase difference as $\Delta\theta_1$.

S3: Calculate the light velocity in the air using the following equation.

$$c = \frac{360 \times d \times f}{|\Delta\theta_1 - \Delta\theta_0|} \quad (1.5)$$

**4.2 Refractive index measurement**

The refractive index measurement of optical transparent material can be achieved based on the optical setup as shown in Fig. 1 in the following process.

S1: Remove the sample off the beam path, and set the frequency of the square waveform emitted through the two output channels of the function generator both to $f$, then use the lock-in amplifier to obtain the phase difference $\Delta\theta_0$ between the reference square signal produced by the function generator and the electrical signal emitted from the high speed photo receiver.

S2: Insert the sample into the beam path, then repeat S1, and write the obtained phase difference as $\Delta\theta_1$.

S3: Calculate the refractive index of the sample using the following equation.

$$n = \frac{|\Delta\theta_1 - \Delta\theta_0| \times c}{360 \times d \times f} + 1 \quad (1.6)$$

Where $c$ is the velocity of light in the air, and $d$ represent the length of the sample in the optical transmission path, $\Delta\theta_0$ and $\Delta\theta_1$ are recorded in the unit of degree.

The refractive index is one of the most important optical parameters of a material, and currently the most popular and reliable technique of determining the sample's refractive index is using the minimum deviation angle [18-20], however within this method, the sample has to be fabricated precisely into a triple prism with a vertex angle for example 60°, which is remarkably time-consuming, and in the measuring procedure, sophisticated mechanical structure has to be employed to obtain the minimum deviation angle by rotating step by step the workbench, the measurement process is thus very tedious and any inappropriate operation would deliver nonnegligible error. On the contrary, potential advantages of our proposed method for determining the refractive index or optical transparent material is the easy preparation of the sample, the convenient operability and the fast measuring speed.

Here it should be pointed out that based on our current hardware configuration, the setup cannot be applied to measure the speed of light, nor the refractive index of transparent material, the main reason is the highest frequency of the square signal the present system can generate and manage is about 20 MHz, which is so low that the phase difference when changing the optical propagation distance or inserting a sample into the beam path is undetectable, we estimate the appropriate frequency for successively determining the velocity of light and the refractive index of ordinary glass is about 1 GHz - 10 GHz, the hardware will get upgraded and our systematic experiment will be carried out on that condition.

## 5. Conclusion

In conclusion, we build a simple setup for measuring the optical propagation time in matter for a known distance, primary experimental result confirms the feasibility of the newly proposed measuring theory, which can be used to measure the speed of light in air and the refractive index of optical transparent material. Compared with the currently similar technique of determining light speed using oscilloscope, potential advantage of our proposed method employing lock-in amplifier is that high accuracy are promising, and in contrast with the presently most popular method for measuring the sample's refractive index based on the minimum deviation angle, superiority of our suggested method is the easy preparation of the sample, the convenient operability and the fast measuring speed.

**Acknowledgment**

This research was partly supported by the National Natural Science Foundation of China (Nos. 61705246 and 11602280) and the scientific equipment developing project of the Chinese academy of sciences (No.28201631231100101).